\magnification=1000
		\catcode`\"=12
	\font\black=cmbx10
\font\sblack=cmbx7
\font\ssblack=cmbx5
\font\blackital=cmmib10  \skewchar\blackital='177
\font\sblackital=cmmib7  \skewchar\sblackital='177
\font\ssblackital=cmmib5  \skewchar\ssblackital='177
%\font\sanss=cmss10
%\font\ssanss=cmss8 scaled 900
%\font\sssanss=cmss8 scaled 600
\font\sanss=txss
\font\ssanss=txss scaled 650
\font\sssanss=txss scaled 500
\font\blackboard=msbm10
\font\sblackboard=msbm7
\font\ssblackboard=msbm5
%\font\caligr=eusm10
%\font\scaligr=eusm7
%\font\sscaligr=eusm5
\font\caligr=zplmr7y
\font\scaligr=zplmr7y scaled 650
\font\sscaligr=zplmr7y scaled 500

\font\fraktur=eufm10
\font\sfraktur=eufm7
\font\ssfraktur=eufm5

\font\amsa=msam10

	\def\all#1{\setbox0=\hbox{\lower1.5pt\hbox{\bsymb
       \char"38}}\setbox1=\hbox{$_{#1}$} \box0\lower2pt\box1\;}
	\def\exi#1{\setbox0=\hbox{\lower1.5pt\hbox{\bsymb \char"39}}
       \setbox1=\hbox{$_{#1}$} \box0\lower2pt\box1\;}

\font\bsymb=cmsy10 scaled\magstep2
\font\CL=cmcsc10

\newfam\bifam
\textfont\bifam=\blackital
\scriptfont\bifam=\sblackital
\scriptscriptfont\bifam=\ssblackital
\def\bi#1{{\fam\bifam\relax#1}}

\newfam\blfam
\textfont\blfam=\black
\scriptfont\blfam=\sblack
\scriptscriptfont\blfam=\ssblack

\newfam\bbfam
\textfont\bbfam=\blackboard
\scriptfont\bbfam=\sblackboard
\scriptscriptfont\bbfam=\ssblackboard
\def\bb#1{{\fam\bbfam\relax#1}}

\newfam\ssfam
\textfont\ssfam=\sanss
\scriptfont\ssfam=\ssanss
\scriptscriptfont\ssfam=\sssanss
\def\ss#1{{\fam\ssfam\relax#1}}

\newfam\clfam
\textfont\clfam=\caligr
\scriptfont\clfam=\scaligr
\scriptscriptfont\clfam=\sscaligr

\newfam\frfam
\textfont\frfam=\fraktur
\scriptfont\frfam=\sfraktur
\scriptscriptfont\frfam=\ssfraktur

\def\hpb#1{\setbox0=\hbox{${#1}$}
    \copy0 \kern-\wd0 \kern.2pt \box0}
\def\vpb#1{\setbox0=\hbox{${#1}$}
    \copy0 \kern-\wd0 \raise.08pt \box0}
\def\pmb#1{\setbox0\hbox{${#1}$} \copy0 \kern-\wd0 \kern.2pt \box0}
\def\pmbb#1{\setbox0\hbox{${#1}$} \copy0 \kern-\wd0
      \kern.2pt \copy0 \kern-\wd0 \kern.2pt \box0}
\def\pmbbb#1{\setbox0\hbox{${#1}$} \copy0 \kern-\wd0
      \kern.2pt \copy0 \kern-\wd0 \kern.2pt
    \copy0 \kern-\wd0 \kern.2pt \box0}
\def\pmxb#1{\setbox0\hbox{${#1}$} \copy0 \kern-\wd0
      \kern.2pt \copy0 \kern-\wd0 \kern.2pt
      \copy0 \kern-\wd0 \kern.2pt \copy0 \kern-\wd0 \kern.2pt \box0}
\def\pmxbb#1{\setbox0\hbox{${#1}$} \copy0 \kern-\wd0 \kern.2pt
      \copy0 \kern-\wd0 \kern.2pt
      \copy0 \kern-\wd0 \kern.2pt \copy0 \kern-\wd0 \kern.2pt
      \copy0 \kern-\wd0 \kern.2pt \box0}

\def\blacktriangle{{\setbox0=\hbox{\amsa\char"4E}\box0}} 
\def\leqslant{{\setbox0=\hbox{\amsa\char"36}\box0}}

\mathchardef\za="710B  %\alpha
\mathchardef\zb="710C  %\beta
\mathchardef\zg="710D  %\gamma
\mathchardef\zd="710E  %\delta
\mathchardef\zve="710F %\epsilon
\mathchardef\zz="7110  %\zeta
\mathchardef\zh="7111  %\eta
\mathchardef\zvy="7112 %\theta
\mathchardef\zi="7113  %\iota
\mathchardef\zk="7114  %\kappa
\mathchardef\zl="7115  %\lambda
\mathchardef\zm="7116  %\mu
\mathchardef\zn="7117  %\nu
\mathchardef\zx="7118  %\xi
\mathchardef\zp="7119  %\pi
\mathchardef\zr="711A  %\rho
\mathchardef\zs="711B  %\sigma
\mathchardef\zt="711C  %\tau
\mathchardef\zu="711D  %\upsilon
\mathchardef\zvf="711E %\phi
\mathchardef\zq="711F  %\chi
\mathchardef\zc="7120  %\psi
\mathchardef\zw="7121  %\omega
\mathchardef\ze="7122  %\varepsilon
\mathchardef\zy="7123  %\vartheta
\mathchardef\zvp="7124 %\varpi
\mathchardef\zvr="7125 %\varrho
\mathchardef\zvs="7126 %\varsigma
\mathchardef\zf="7127  %\varphi
\mathchardef\zG="7000  %\Gamma
\mathchardef\zD="7001  %\Delta
\mathchardef\zY="7002  %\Theta
\mathchardef\zL="7003  %\Lambda
\mathchardef\zX="7004  %\Xi
\mathchardef\zP="7005  %\Pi
\mathchardef\zS="7006  %\Sigma
\mathchardef\zU="7007  %\Upsilon
\mathchardef\zF="7008  %\Phi
\mathchardef\zC="7009  %\Psi
\mathchardef\zW="700A  %\Omega

	\font\kropa=lcircle10 scaled 1700
	\def\tdot{\setbox0=\hbox{\kropa \char"70} \kern1.5pt \raise.35pt \box0}
	\def\sdot{\setbox0=\hbox{\kropa \char"70} \kern1.5pt \raise2.9pt \box0}

	\def\C{{\bb C}}
	\def\R{{\bb R}}

	\def\sT{{\ss T}}

	\def\bI{{\bi I}}
	\def\bM{{\bi M}}
	\def\bN{{\bi N}}

	\def\bT{{\bi T}}

	\catcode`\"=\active

	\newcount\secnum  \secnum=0 
	\newcount\subsecnum
	\newcount\subsubsecnum 

	\newcount\Asecnum  \secnum=0 
	\newcount\Asubsecnum

	\newcount\tcount  \tcount=0
	\newcount\pcount  \pcount=0
	\newcount\dcount  \dcount=0
	\newcount\ecount  \ecount=0

	\def\Theorem#1{\global\advance\tcount by 1
			\vskip2pt
			\noindent{\CL Theorem \the\tcount.} {\it #1}\par\vskip2pt}

	\def\Proposition#1{\global\advance\pcount by 1
			\vskip2pt
			\noindent{\CL Proposition \the\pcount.} {\it #1}\par\vskip2pt}

	\def\Definition#1{\global\advance\dcount by 1
			\vskip2pt
			\noindent{\CL Definition \the\dcount.} #1
			{\hfill \blacktriangle}\par\vskip2pt}

	\def\Example#1{\global\advance\ecount by 1
			\vskip2pt
			\noindent{\CL Example \the\ecount.} #1
			{\hfill \blacktriangle}\par\vskip2pt}

	\def\sect#1{ 
            \global\advance \secnum by 1 \subsecnum=0 \subsubsecnum=0 
			\vskip2pt
            \noindent{\bf \the\secnum.\ #1 \vskip-4.5mm}
			\nobreak
			\leftline{}}

	\def\ssect#1{
	    	\global\advance\subsecnum by 1 \subsubsecnum=0
			\vskip1pt
            \noindent{\bf \the \secnum.\the\subsecnum.\ #1 \vskip-4.5mm}
		    \nobreak
			\leftline{}}

	\def\sssect#1{\global\advance\subsubsecnum by 1 
            \vskip1pt
			\noindent{\bf \the\secnum.\the\subsecnum.\the\subsubsecnum.\ #1 \vskip-4.5mm} 
	        \nobreak
		    \leftline{}}

	\def\Asect#1{ 
            \global\advance \Asecnum by 1 \Asubsecnum=0 %\subsubsecnum=0 
			\vskip2pt
            \noindent{\bf \the \Asecnum.\ #1 \vskip-4.5mm}
			\nobreak
			\leftline{}}

	\def\Assect#1{
	    	\global\advance\Asubsecnum by 1 %\subsubsecnum=0
			\vskip1pt
            \noindent{\bf A\the \Asecnum .\the\Asubsecnum\  #1 \vskip-4.5mm}
		    \nobreak
			\leftline{}}

	\font\Tfont=cmb10 scaled 1300

	\font\tfont=cmb10 %scaled 1300

	\def\Title#1{\vskip2mm\centerline{\Tfont #1}\vskip1.5mm}

	\def\title#1{\vskip1mm\leftline{\tfont #1}}%\vskip.5mm}

	\def\compose#1#2#3#4#5#6{{\setbox0=\hbox{\raise#2\hbox{\kern#3\hbox{${#1}$}}}
\setbox1=\hbox{\raise#5\hbox{\kern#6\hbox{${#4}$}}}\box0\box1}}

	\def\position#1#2#3{{\setbox0=\hbox{\raise#2\hbox{\kern#3\hbox{${#1}$}}}\box0}}

	\def\Sim#1{\kern.5pt\setbox0=\hbox{$\VPU{8pt}^{\scriptscriptstyle{#1}}$}\setbox1=\hbox{$\sim$}
			\setbox2=\hbox{\kern.5\wd0\copy1\kern-.5\wd0\kern-3pt\copy0}\box2\kern2.7pt}

	\def\RF{\null}

	\def\VPU#1{{\vrule height#1 width0pt depth0pt}}
	\def\VPD#1{{\vrule height0pt width0pt depth#1}}

	\def\lpr{{\setbox0=\hbox{\vrule height .15pt width 3.5pt depth 0pt}
\setbox1=\hbox{\vrule height 5.8pt width .3pt depth 0pt}\kern2pt\box0\box1\kern3pt}}

	\def\rpr{{\setbox0=\hbox{\vrule height .15pt width 3.5pt depth 0pt}
\setbox1=\hbox{\vrule height 5.8pt width .3pt depth 0pt}\kern3pt\box0\kern-3.5pt\box1\kern6.5pt}}

	\def\fpr#1{\kern-4pt\setbox0=\hbox{$\VPD{7pt}_{#1}$}\setbox1=\hbox{$\times$}
			\setbox2=\hbox{\kern.5\wd0\copy1\kern-.5\wd0\kern-3pt\copy0}\box2\kern0pt}

	\def\*{{\textstyle *}}

\def\*{{\textstyle *}}
\def\s*{{\scriptstyle *}}

	\def\leqs{\kern3pt\leqslant\kern3pt}

%	\def\all#1{\setbox0=\hbox{\lower1.5pt\hbox{\bsymb
%       \char"38}}\setbox1=\hbox{$_{#1}$} \box0\lower2pt\box1\;}
%	\def\exi#1{\setbox0=\hbox{\lower1.5pt\hbox{\bsymb \char"39}}
%       \setbox1=\hbox{$_{#1}$} \box0\lower2pt\box1\;}

	%File information: C:\PTEXTS\AEDIT\MACRO.TEX X     128    6/11/07   10:56
%--------------------------------------------------------------

%\def\fpr#1{\kern-4pt\setbox0=\hbox{$\VPD{7pt}_{#1}$}\setbox1=\hbox{$\times$}
%			\setbox2=\hbox{\kern.5\wd0\copy1\kern-.5\wd0\kern-3pt\copy0}\box2\kern0pt}

\def\osT{\sT \position{\scriptscriptstyle\circ}{9pt}{-5.5pt}\kern3pt}

\def\WMT{
		\centerline{W\l odzimierz M. Tulczyjew }
		\centerline{Valle San Benedetto, 2 }
        \centerline{62030 Monte Cavallo, Italy }
		\centerline{Associated with }
        \centerline{Division of Mathematical Methods in Physics}
        \centerline{University of Warsaw}
        \centerline{Ho\.{z}a 74, 00-682 Warszawa}
	\centerline{and}
        \centerline{Istituto Nazionale di Fisica Nucleare,}
        \centerline{Sezione di Napoli}
        \centerline{Complesso Universitario di Monte Sant'Angelo}
        \centerline{Via Cinthia, 80126 Napoli, Italy}
		\centerline{{\tt tulczy@libero.it} }
		\vskip1mm
		}

\catcode`\"=12

\catcode`\"=\active

\def\text#1{{\rm\hskip2mm #1 \hskip2mm}}

%Label statistics: 0	0	Bad references: 0
	\hsize=37truepc
	\hoffset=-10truept
	\vsize=52truepc
	\voffset=6truept
	
	\input pictex
	\input DCpic.sty

	\input xy
	\xyoption{all}

\vskip5mm

		\Title{Density operators and selective measurements.}

		\WMT

		\sect{Introduction.}
	It is widely believed that statistical interpretation of quantum mechanics requires that density operators representing
quantum states be normalized.  We present a description of selective measurements in terms of density operators.  The
description is inspired by Schwinger's Algebra of Microscopic Measurements [1], (see also [2]).  Density operators used are
not normalized.  We do not know applications of density operators requiring normalization.

		\sect{Beams of particles.}
	The physical space is an affine space $M$ of dimension 3 modelled on a vector space $V$.  There is an Euclidean
metric tensor
		$$g \,\colon V \rightarrow V^\*.
																												\eqno(1)$$

	We consider beams of particles of mass $m$ and constant energy $E$ in the direction of a unit vector $z \in V$. The
internal states of the particles are elements of a unitary vector space $U$ of dimension $r$ over the field $\C$ of complex
numbers.  The elements of $U$ are {\it kets} $|u\rangle$ and elements of the dual space $U^\*$ are {\it bras} $\langle a|$.
The unitary structure establishes an antilinear isomorphism of $U$ with $U^\*$ assigning to each ket $|u\rangle$ a unique
bra $\langle u^\dagger|$.  The number
		$$\langle u_1{}^\dagger|u_2\rangle
																												\eqno(2)$$
	is the {\it scalar product} of vectors $|u_1\rangle$ and $|u_2\rangle$

	We introduce a sequence of points
		$$x_0, x_1, x_2, \ldots , x_n
																												\eqno(3)$$
	satifying inequalities
		$$\langle g(z), x_i - x_{i-1}\rangle > 0,
																												\eqno(4)$$
	and a corresponding sequence of planes
		$$X_i = \left\{x \in M ;\; \langle g(z), x - x_i\rangle = 0 \right\}.
																												\eqno(5)$$
	In the immediate neighbourhood of each plane $X_i$ the beam is not subject to external interaction and is represented
by a plane wave
		$$|\zc_i(x)\rangle = |A_i\rangle \exp\left(ik\langle g(z), x - x_0\rangle + \zd_i\right),\;\;\; |A_i\rangle \in
U,\;\;\; k = {{\sqrt{2mE}}\over{\hbar}}
																												\eqno(6)$$
	with time dependence separated.  The probability flux through a unit surface element of the plane $X_i$ is expressed by
		$${{\hbar k}\over{m}} \langle A_i{}^\dagger|A_i\rangle .
																												\eqno(7)$$

	Between the plane $X_{i-1}$ and the plane $X_i$ the beam passes through a selective device.  Its action on the wave
function is described as the action of a linear transition operator
		$$M_i \,\colon U \rightarrow U \,\colon |\zc_{i-1}(x)\rangle \mapsto |\zc_i(x)\rangle = M_i |\zc_{i-1}(x)\rangle.
																												\eqno(8)$$
	If
		$$M_1, M_2, M_3, \ldots , M_n
																												\eqno(9)$$
	is the sequence of transition operators and
		$$|\zc_0(x)\rangle = |A_0\rangle \exp\left(ik\langle g(z), x - x_0\rangle\right)
																												\eqno(10)$$
	is the initial state, then
		$$\eqalign{
		|\zc_i(x)\rangle &= |A_i \rangle \exp\left(ik\langle g(z), x - x_0\rangle + \zd_i\right) \cr
	&= M_i \cdots M_2 M_1 |A_0\rangle \exp\left(ik\langle g(z), x - x_0\rangle\right)}
																												\eqno(11)$$
	and
		$$\langle \zc_i{}^\dagger(x)| \zc_i(x)\rangle = \langle A_0{}^\dagger|M_1{}^\dagger M_2{}^\dagger \cdots
M_i{}^\dagger M_i \cdots M_2 M_1 |A_0 \rangle
																												\eqno(12)$$
	The flux of particles through unit surface element of $X_i$ is given by
		$${{\hbar k}\over{m}} \langle A_0{}^\dagger|M_1{}^\dagger M_2{}^\dagger \cdots
M_i{}^\dagger M_i \cdots M_2 M_1 |A_0 \rangle.
																												\eqno(13)$$

		\sect{Mixed states and density operators.}
	The expression \RF(13) can be presented in the form
		$${{\hbar k}\over{m}} \;{\rm tr}\left(M_i \cdots M_2 M_1 |A_0
\rangle\langle A_0{}^\dagger| M_1{}^\dagger M_2{}^\dagger \cdots M_i{}^\dagger \right).
																												\eqno(14)$$
	In this new expression the pure initial state is represented by the density operator $|A_0 \rangle\langle
A_0{}^\dagger|$ and can be replaced by a mixed state represented by a positive Hermitian density operator $\bT$.  The
expression
		$${{\hbar k}\over{m}} \;{\rm tr}\left(M_i \cdots M_2 M_1 \bT M_1{}^\dagger M_2{}^\dagger \cdots M_i{}^\dagger
\right)
																												\eqno(15)$$
	is the result.

		\sect{Selective measurements.}
	We set the density operator $\bT$ in the expression \RF(15) equal to
		$$\bT = {{m}\over{\hbar kr}} \bI,
																												\eqno(16)$$
	where $\bI$ is the identity operator.  The operator $\bT$ is normalized in the sense that
		$${{\hbar k}\over{m}}\;{\rm tr}\;\bT = 1
																												\eqno(17)$$
	although this normalization is of no importance.  The expression
		$$P_i = {{\hbar k}\over{m}} \;{\rm tr}\left(M_i \cdots M_2 M_1 \bT M_1{}^\dagger M_2{}^\dagger \cdots M_i{}^\dagger
\right) = {{1}\over{r}} \;{\rm tr}\left(M_i \cdots M_2 M_1 M_1{}^\dagger M_2{}^\dagger \cdots M_i{}^\dagger \right)
																												\eqno(18)$$
	represents the probability of detecting a particle crossing a unit surface element of the plane $X_i$ in unit time.  The
state of the initial beam emitted by a source at $X_0$ is totally mixed.

	We want to describe the following experimental arrangement.  The initial beam emitted at $X_0$ undergoes a preliminary
selection by a sequence of devices represented by the sequence of operators
		$$M_1, M_2, \ldots , M_j.
																												\eqno(19)$$
	The beam undergoes further selection passing through a sequence of devices represented by operators
		$$N_1 = M_{j+1}, N_2 = M_{j+2}, \ldots , N_{n-j} = M_n.
																												\eqno(20)$$
	The particles are detected at $X_n$ by a non selective detector.  After the preliminary selection the state of the beam
is represented by the density operator
		$$\bM = M_j M_{j-1} \cdots M_1 \bT M_1{}^\dagger \cdots M_{j-1}{}^\dagger M_j{}^\dagger = {{m}\over{\hbar kr}} M_j
M_{j-1} \cdots M_1 M_1{}^\dagger \cdots M_{j-1}{}^\dagger M_j{}^\dagger
																												\eqno(21)$$
	with the probability of non selective detection
		$$P_{in} = {{\hbar k}\over{m}} \;{\rm tr}\left(M_j M_{j-1} \cdots M_1 \bT M_1{}^\dagger \cdots M_{j-1}{}^\dagger
M_j{}^\dagger \right) = {{\hbar k}\over{m}} \;{\rm tr}\;\bM
																												\eqno(22)$$
	The beam arrives at $X_n$ in a state represented by the density operator
		$$N_{n-j} N_{n-j-1} \cdots N_1 M_j M_{j-1} \cdots M_1 \bT M_1{}^\dagger \cdots M_{j-1}{}^\dagger M_j{}^\dagger
N_1{}^\dagger \cdots N_{n-j-1}{}^\dagger N_{n-j}{}^\dagger
																												\eqno(23)$$
	It is detected with the probability
		$$\eqalign{
			P_{out} &= {{\hbar k}\over{m}} \;{\rm tr}\left(N_{n-j} N_{n-j-1} \cdots N_1 M_j M_{j-1} \cdots M_1 \bT
M_1{}^\dagger \cdots M_{j-1}{}^\dagger M_j{}^\dagger N_1{}^\dagger \cdots N_{n-j-1}{}^\dagger N_{n-j}{}^\dagger \right) \cr
		&= {{\hbar k}\over{m}} \;{\rm tr}\left(N_1{}^\dagger \cdots N_{n-j-1}{}^\dagger N_{n-j}{}^\dagger N_{n-j} N_{n-j-1}
\cdots N_1 M_j M_{j-1} \cdots M_1 \bT M_1{}^\dagger \cdots M_{j-1}{}^\dagger M_j{}^\dagger \right) \cr
		&= {{\hbar k}\over{m}} \;{\rm tr}(\bN\bM)}
																												\eqno(24)$$
	with
		$$\bN = N_1{}^\dagger \cdots N_{n-j-1}{}^\dagger N_{n-j}{}^\dagger N_{n-j} N_{n-j-1} \cdots N_1.
																												\eqno(25)$$
	The density operator $\bN$ characterizes the selective detector.  The probability $P_{out}$ is measured at $X_j$ by the
selective detector.  This measurement is performed on the mixed state represented by the operator $\bM$.  The relative
probability
		$$P_{out}/P_{in} = {\rm tr}(\bN\bM)/{\rm tr}\,\bM
																												\eqno(26)$$
	should be considered the result of the selectve measurement described.  Arbitrary normalization can be imposed on
$\bM$.  Normalization of $\bN$ would distort the result of the measurement.

		\sect{An example.}
	In addition to the metric tensor
		$$g \,\colon V \rightarrow V^\*
																												\eqno(27)$$
	we introduce in the model space $V$ of the physical space $M$ an orientation $o$ defined as an equivalence class of
bases.

	We analyse the internal states of a beam of particles of spin $1/2$.  States of particles are represented by wave
functions with values in an unitary space of complex dimension 2.  The set of hermitian traceless operators in $U$ is a
real vector space $S$ of dimension 3.

	The trace ${\rm tr}\,(ab)$ of a product is a non negative real number and the mapping
		$$S \times S \rightarrow \R\,\colon (a,b) \mapsto {\rm tr}\,(ab)
																												\eqno(28)$$
	is bilinear and symmetric.  The spectrum of an operator $a \in S$ is a pair $\{\za,-\za\}$ of real numbers and the
spectrum of the operator $aa$ is the set $\{\za^2,\za^2\}$.  It follows that ${\rm tr}\,(aa) = 0$ if and only if $a = 0$.
In conclusion we have a Euclidean {\it scalar product}
		$$(\,\;|\,\;) \,\colon S \times S \rightarrow \R\,\colon (a,b) \mapsto (a|b) = {{1}\over{2}} {\rm tr}\,(ab).
																												\eqno(29)$$

	We introduce the {\it Pauli morphism}
		$$\zs \,\colon V \rightarrow S.
																												\eqno(30)$$
	This morphism is an isometry such that the operator
		$${{1}\over{2}} \zs(w) \,\colon U \rightarrow U
																												\eqno(31)$$
	associated with each unit vector $w \in V$ is the {\it spin operator} in the direction $w$.  Its spectrum is the set
$\{1/2,-1/2\}$ and its eigenvectors represent states of the particle with spin $1/2$ and $-1/2$ in the direction of $w$.

	We introduce a number of operators in the space $U$:
	\item{ 1)} The projection operator
		$$K(w) = {{1}\over{2}}\left(I + \zs(w)\right)
																												\eqno(32)$$
	associated with a unit vector $w \in V$.  This operator projects onto the space of eigenstates of the spin operator
$1/2\zs(w)$ corresponding to the eigenvalue $1/2$.

	\item{ 2)} A {\it phase shift} operator
		$$D(\zd) = \exp(i\zd) I.
																												\eqno(33)$$

	\item{ 3)} An {\it attenuation} operator
		$$R(\zr) = \exp(-\zr/2) I.
																												\eqno(34)$$
	\item{ 4)} A unitary unimodular operator
		$$G \,\colon U \rightarrow U.
																												\eqno(35)$$
	This operator represents a rotation
		$$E \,\colon V \rightarrow V
																												\eqno(36)$$
	in the sense that
		$$G \zs(w) G^{-1} = \zs(Ew).
																												\eqno(37)$$

	\item{ 5)} The operator
		$$Q(w) = {{1}\over{2}}\left(I + \zs(w)\right)
																												\eqno(38)$$
	associated with a vector $w \in V$ of norm $\|w\| \neq 1$.  This operator is not a projection operator.

	Consider a beam undergo a preliminary selection by devices represented by $M_1 = K(w)$ and $M_2 = D(\zd)$.  The vector
$w$ is orthogonal to the direction of the beam and the first of the devices is a Stern-Gerlach filter.  It is accompanied
by an unavoidable phase shift.  The state prepared by these devices is a pure state represented by the density operator
		$$\bM = M_2 M_1 T M_1{}^\dagger M_2{}^\dagger = {{m}\over{2\hbar k}} K(w).
																												\eqno(39)$$
	The selective detector is composed of devices represented by operators $N_1 = R(\zr)$, $N_2 = D(\zd')$, and $N_3 =
K(w')$.  The attenuation $R(\zr)$ my be due to the beam passing through a potential barrier.  The density operator
		$$\bN = N_1{}^\dagger N_2{}^\dagger N_3{}^\dagger N_3 N_2 N_1 = \exp(-\zr)K(w')
																												\eqno(40)$$
	represents the selective detector.  The result of the selective measurement is the relative probability
		$$P_{out}/P_{in} = {\rm tr}(\bN\bM)/{\rm tr}\,\bM = {{1}\over{2}} \left(1 + (w'|w)\right)
																												\eqno(41)$$
	since
		$${\rm tr}\,\bM = 1
																												\eqno(42)$$
	and
		$$\eqalign{
			{\rm tr}(\bN\bM) &= {\rm tr}\left(\exp(-\zr)K(w')K(w)\right) \cr
		&= {{1}\over{4}} \,{\rm tr}\left(\exp(-\zr)(I + \zs(w'))(I + \zs(w))\right) \cr
		&= {{1}\over{4}} \,{\rm tr}\left(\exp(-\zr)(I + \zs(w') + \zs(w) + \zs(w')\zs(w))\right) \cr
		&= {{1}\over{2}} \left(1 + (w'|w)\right)    }
																												\eqno(43)$$

		\sect{References.}
\vskip1mm
\noindent [1] J. Schwinger, {\it The Algebra of Microscopic Measurements}, Proc. Natl. Acad. Sc. US, {\bf 45} (1959)

\vskip1mm
\noindent [2] F. A. Kaempffer, {\it Concepts in Quantum Mechanics}, Academic Press, New York and London (1965)

\end